\documentclass[useAMS,usenatbib,usegraphicx,letterpaper]{mn2e}

\newcommand{\be}{\begin{equation}}
\newcommand{\ee}{\end{equation}}
\newcommand{\ba}{\begin{eqnarray}}
\newcommand{\ea}{\end{eqnarray}}

\def\simless{\mathbin{\lower 3pt\hbox
   {$\rlap{\raise 4pt\hbox{$\char'074$}}\mathchar"7218$}}}
\def\simgreat{\mathbin{\lower 3pt\hbox
   {$\rlap{\raise 4pt\hbox{$\char'076$}}\mathchar"7218$}}}   % > or of order

\title[The UKIDSS DR1]{The UKIRT Infrared Deep Sky Survey First Data
Release} \author[S. J. Warren, et al.]{S. J. Warren$^1$\thanks{E-mail:
s.j.warren@imperial.ac.uk}, 
N. C. Hambly$^2$, 
S. Dye$^3$,
O. Almaini$^4$, 
N. J. G. Cross$^2$,
A. C. Edge$^5$, \newauthor
S. Foucaud$^4$, 
P. C. Hewett$^6$, 
S. T. Hodgkin$^6$,
M. J. Irwin$^6$,
R. F. Jameson$^7$, \newauthor  
A. Lawrence$^2$, 
P. W. Lucas$^8$,
A. J. Adamson$^9$, 
R. M. Bandyopadhyay$^{10}$,
J. Bryant$^2$,  \newauthor 
R. S. Collins$^2$, 
C. J. Davis$^9$,
J. S. Dunlop$^2$,
J. P. Emerson$^{11}$
D. W. Evans$^6$, \newauthor
E. A. Gonzales-Solares$^6$,
P. Hirst$^9$,
M. J. Jarvis$^{12}$, 
T. R. Kendall$^8$,
T. H. Kerr$^9$, \newauthor
S. K. Leggett$^9$,
J. R. Lewis$^6$,  
R. G. Mann$^2$,
R. J. McLure$^2$,
R. G. McMahon$^6$, \newauthor 
D. J. Mortlock$^1$,
M. G. Rawlings$^9$,
M. A. Read$^2$,
M. Riello$^6$,
C. Simpson$^{13}$, \newauthor
D. J. B. Smith$^{12}$,
E. T. W. Sutorius$^2$,
T. A. Targett$^2$,
W. P. Varricatt$^9$
\vspace{7mm}\\
$^1$Astrophysics Group, Imperial College London, Blackett Laboratory,
Prince Consort Road, London, SW7 2AZ, U.K.\\
$^2$Scottish Universities Physics Alliance, Institute for Astronomy, 
University of Edinburgh, \\
Royal Observatory Edinburgh, Blackford Hill, Edinburgh, EH9 3HJ, U.K.\\
$^3$Cardiff University, School of Physics \& Astronomy, Queens Buildings,
The Parade, Cardiff, CF24 3AA, U.K. \\
$^4$School of Physics and Astronomy, University of Nottingham,
University Park, Nottingham, NG7 2RD, U.K. \\ 
$^5$Department of Physics, Durham University, South Road, DH1 3LE,
U.K. \\
$^6$Institute of Astronomy, Madingley Rd., Cambridge, CB3 0HA, U.K.\\
$^7$Department of Physics and Astronomy, University of Leicester,
Leicester, LE1 7RH, U.K. \\
$^8$Centre for Astrophysics Research, Science and Technology Research 
Institute, University of Hertfordshire, Hatfield, AL10 9AB, U.K. \\
$^9$Joint Astronomy Centre, 660 N. A'ohoku Place, University Park,
Hilo, Hawaii 96720, U.S.A.\\
$^{10}$Department of Astronomy, University of Florida, 211 Bryant Space
Science Center, Gainesville, Florida 32611, U.S.A.\\
$^{11}$Astronomy Unit, School of Mathematical Sciences, Queen Mary,
University of London, Mile End Road, London E1 4NS, U.K. \\
$^{12}$Department of Physics,  Denys Wilkinson Building, Keble Road,
Oxford, OX1 3RH, U.K.  \\
$^{13}$Astrophysics Research Institute, Liverpool John Moores University,
Twelve Quays House, Egerton Wharf, Birkenhead, CH41 1LD, U.K.\\}

\begin{document}

\date{}

\pagerange{\pageref{firstpage}--\pageref{lastpage}} \pubyear{2006}

\maketitle

\label{firstpage}

\begin{abstract}
The First Data Release (DR1) of the UKIRT Infrared Deep Sky Survey
(UKIDSS) took place on 2006 July 21. UKIDSS is a set of five large
near--infrared surveys, covering a complementary range of areas,
depths, and Galactic latitudes.  DR1 is the first large release of
survey-quality data from UKIDSS and includes 320 deg$^2$ of
multicolour data to (Vega) $K=18$, complete (depending on the survey) in
three to five bands from the set {\em ZYJHK}, together with 4 deg$^2$
of deep {\em JK} data to an average depth $K=21$. In addition the
release includes a similar quantity of data with incomplete filter
coverage.  In {\em JHK}, in regions of low extinction, the photometric
uniformity of the calibration is better than 0.02\,mag. in each
band. The accuracy of the calibration in {\em ZY} remains to be
quantified, and the same is true of {\em JHK} in regions of high
extinction. The median image FWHM across the dataset is
$0.82\arcsec$. We describe changes since the Early Data Release in the
implementation, pipeline and calibration, quality control, and archive
procedures. We provide maps of the areas surveyed, and summarise the
contents of each of the five surveys in terms of filters, areas, and
depths. DR1 marks completion of 7 per cent of the UKIDSS 7-year goals.

\end{abstract}

\begin{keywords}
astronomical data bases: surveys -- infrared: general
\end{keywords}

\section{Introduction}

UKIDSS is the UKIRT Infrared Deep Sky Survey \citep{lawrence06},
carried out using the Wide Field Camera \citep[WFCAM;][]{casali06}
installed on the United Kingdom Infrared Telescope (UKIRT). Data
acquisition for the survey started in 2005 May. A prototype dataset,
the Early Data Release (EDR), was released on 2006 February 10, and is
described in \citet{dye06} (hereafter D06). The present paper defines
the UKIDSS First Data Release (DR1), the first large release of UKIDSS
survey-quality data. The data were released to the ESO community on
2006 July 21, and are available from {\tt
  http://surveys.roe.ac.uk/wsa}.
\footnote{World release of UKIDSS data products follows after an 18
month interval.}

UKIDSS is a programme of five imaging surveys that each uses some or
all of the broadband filter complement $ZYJHK$, and that span a range
of areas, depths, and Galactic latitudes. There are three high
Galactic latitude surveys, providing complementary combinations of
area and depth; the Large Area Survey (LAS), will cover 4000 deg$^2$
to $K=18$, the Deep ExtraGalactic Survey (DXS), 35 deg$^2$ to
$K=21$, and the Ultra Deep Survey (UDS), 0.8 deg$^2$ to $K=23$.  There
are two other wide surveys to $K=18$, aimed at targets in the Milky
Way; the Galactic Plane Survey (GPS) will cover 1900 deg$^2$, and the
Galactic Clusters Survey (GCS) 1100 deg$^2$. The complete UKIDSS
programme is scheduled to take seven years, requiring $\sim$1000
nights on UKIRT. The current implementation strategy is focused on
completing an intermediate set of goals, defined by the `2-year plan'
detailed in D06. All magnitudes quoted in this paper use the Vega
system described by \citet{hewett06}. Depths, where not explicitly
specified, are the total brightness of a point source for which the
flux integrated in a $2\arcsec$ diameter aperture is detected at
$5\sigma$.

A set of five baseline papers provides the relevant technical
background information for the surveys. The overview of the programme
is given by \citet{lawrence06}. This sets out the science goals that
drove the design of the survey programme, and details the final
coverage that will be achieved, in terms of fields, areas, depths, and
filters. The camera is described in detail by \citet{casali06}, and
the {\em ZYJHK} photometric system is characterised by
\citet{hewett06}, who provide synthetic colours for a wide range of
types of star, galaxy, and quasar. Details of the data pipeline and
data archive will appear in Irwin et al. (2007, in prep.) and Hambly
et al. (2007, in prep.), respectively.

The EDR (D06), cited earlier, contains a sample of the data obtained
in 2005, and amounts to about one per cent of the 7-year plan. The EDR
represented a step towards regular release of survey-quality data, and
occurred while the quality-control (QC) procedures were still being
developed, and the pipeline finalised.  The EDR included all data from
the first observing block, 05A (2005 April to June), and the DXS and
UDS data up to the end of September in block 05B (2005 August to 2006
January). The 05B dataset is substantially larger, and was processed
with a later version of the pipeline.  The current release, DR1,
includes all the 05A and 05B data. All the 05B data have been verified
using the revised QC procedures that define `survey quality'.  The 05A
data, however, are included in DR1, almost unchanged from the EDR (the
minor changes are detailed in section 3.5) i.e. with the older
versions of the pipeline and QC procedures.\footnote{It had been
  intended to release the reprocessed 05A data in DR1, but this is now
  postponed to the next release.}  In a similar manner to the EDR, two
databases are provided in this release. The DR1 database includes all
data where the full filter complement for a particular survey exists,
and the DR1+ database (which is a superset of the DR1 database)
includes all data that have passed QC. The next observing block, 06A,
2006 May to July, will be combined with the 05A and 05B data, and
released in DR2, early in 2007.

Each UKIDSS data release will be accompanied by a paper summarising
the contents of the release, and detailing procedural changes since
the previous release. The EDR paper, D06, is a self-contained summary
of all information relevant to understanding the contents of the
EDR. It serves as the baseline paper for technical details for all
releases, and in the current paper only changes in technical details
are documented. The terms EDR and DR1 have been copied from the Sloan
Digital Sky Survey (SDSS), and the aims of the UKIDSS EDR and DR1
papers are similar to those of the SDSS EDR \citep{stoughton02} and
DR1 \citep{abazajian03} papers. Besides providing details of the
contents of the EDR database, D06 includes relevant details of the
camera design, the observational implementation (integration times,
microstepping), the pipeline and calibration, data artifacts, and the
quality control procedures, as well as a brief guide to querying the
archive.

In the current paper, in Section \ref{projects} we illustrate the
fields surveyed on an all-sky map. In Section \ref{update} we detail
differences in the implementation, pipeline and calibration, quality
control, and archive procedures between EDR and DR1. In Section
\ref{dr1} we provide maps for each survey, illustrating the coverage
of the DR1 release.  In Section \ref{summary} we summarise the
contents of DR1 in terms of areas and depths, and other relevant
quantities.

\begin{figure*}
\includegraphics[width=17cm]{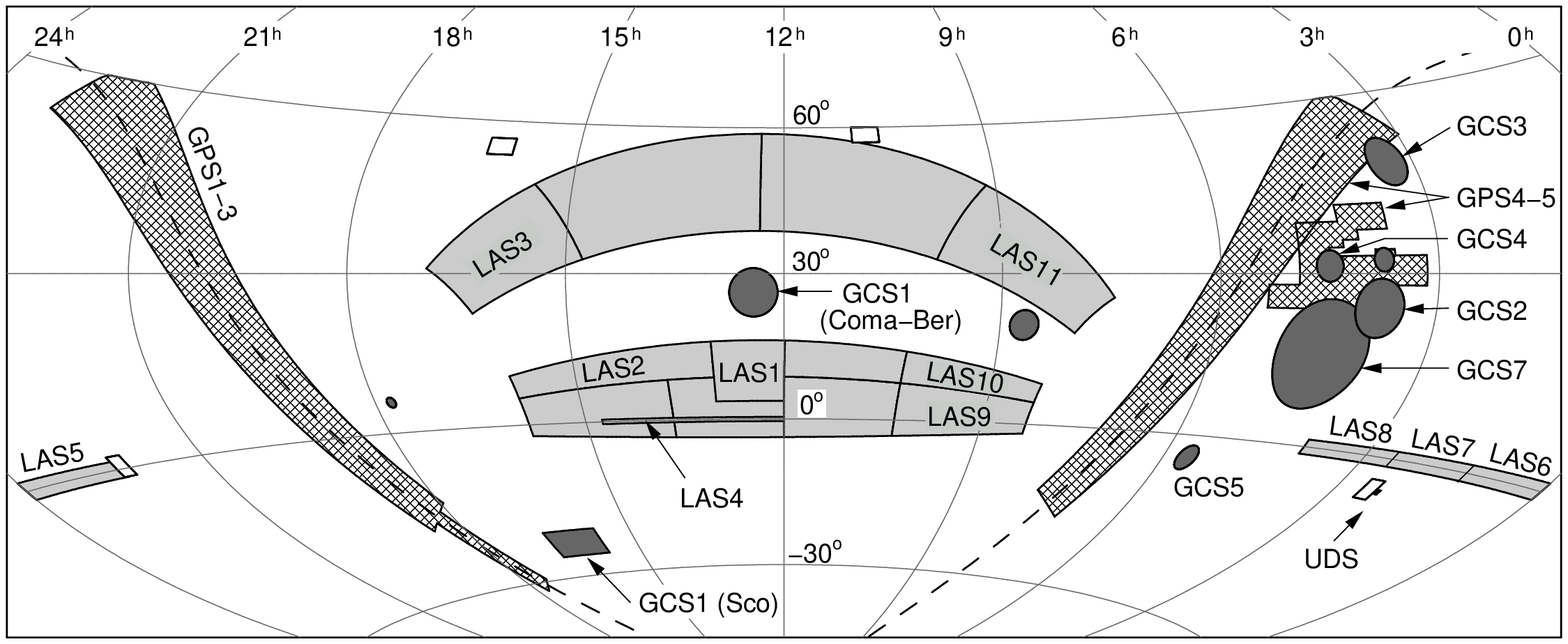}
\caption{The UKIDSS 2--year plan survey areas showing the LAS (solid
  light grey), GPS (cross-hatched), GCS (solid dark grey), DXS (empty
  squares) and UDS (lying alongside the western-most DXS field). The
  dashed line indicates the Galactic plane. LAS, GPS, GCS projects
  included in DR1 are marked. The names of the GCS clusters not labeled
  are Pleiades (GCS2), Alpha Per (GCS3), Tau.--Aur. (GCS4), Orion
  (GCS5), and Hyades (GCS7).}
\label{ukidss_survey_areas_dr1}
\end{figure*}

\section{Overview of DR1 fields}
\label{projects}

In this section we provide an overview of the areas surveyed in
DR1. Detailed coverage maps are provided in Section \ref{dr1}.
Fig. \ref{ukidss_survey_areas_dr1} illustrates the areas that will be
covered in the 2-year plan. The LAS areas are shaded light grey, the
GPS areas are cross-hatched, and the GCS areas are shaded dark grey.
The DXS targets are the four white rectangles with centres as follows;
XMM-LSS (2$^{\rm h}25^{\rm m}$,$-4^{\circ}30'$), the Lockman Hole
(10$^{\rm h}57^{\rm m}$,+57$^{\circ}40'$), Elais N1 (16$^{\rm
  h}10^{\rm m}$,+54$^{\circ}00'$), and VIMOS 4 (22$^{\rm h}17^{\rm
  m}$,+0$^{\circ}20'$). The UDS field covers a single tile,
0.8deg$^2$, and is just visible to the W of the DXS XMM-LSS field.

For administrative purposes the surveys are divided into
`projects'. The LAS, GPS, and GCS projects which contain observations
released in DR1 are illustrated in the figure. In the LAS, the data
from projects LAS$1-4$ were observed in 05A, and were released in the
EDR, and are included in DR1 unchanged, with one proviso. This is that
the algorithm for associating detector frames (Section \ref{archivesec}), in
merging sources across bands, has been refined. For some sources, in
regions where frames overlap, a different choice of frame has
resulted in revised photometry. New projects LAS$5-11$ were observed
in 05B. The GPS is divided into two main regions. The eastern wing
lies mostly between $18^{\rm h}$ and $23^{\rm h}$ and was observed in 05A, in
project GPS1. These data were released in the EDR, and are included
here unchanged (with the above proviso). The eastern wing was also
observed in 05B, in projects GPS2 and 3. The western wing covers the
Galactic plane between approximately $3^{\rm h}$ and $8^{\rm h}$, and an
additional region below the plane, the Taurus--Auriga--Perseus
molecular cloud complex. The western wing was observed in 05B in
projects GPS4 and 5. In the GCS, the targets Sco and Coma--Ber were
observed in 05A in project GCS1. These data were released in the EDR,
and are included here unchanged. In 05B the following targets were
observed: Pleiades, Alpha Per, Tau.--Aur., Orion, and
Hyades. Coordinates for the complete list of GCS targets are provided
in Table 3 in D06.

In the DXS, the Lockman Hole was observed only in 05A, Elais N1 and
XMM-LSS were observed in 05A and 05B, and VIMOS4 was observed only in
05B. The Lockman Hole data, and data taken in Elais N1 and XMM--LSS up
to 2005 Sept 27, were released in the EDR. These data are included in
DR1, with minor differences. Firstly the QC for the 05B data has been
redone, together with the (later) 05B data not included in the EDR,
using the revised procedures detailed in Section
\ref{update}. Secondly, in all fields the stacking has been redone for
DR1, having fixed two problems that were evident in the EDR stacking
(D06). The UDS was only observed in 05B, and has been treated in the
same way i.e. the QC for all the data released in DR1 was done in a
homogeneous manner following the revised procedures.

\section{Update}
\label{update}

D06 contains details of the implementation, pipeline, calibration,
quality control, and archive procedures applied to the EDR data, and
a glossary of technical terms. In this section we describe the
differences between the procedures applied to the 05B data for DR1,
and those applied to the 05A data for the EDR.

\begin{table}
\centering
\begin{tabular}{@{}lccccc@{}}
\hline
Scheme  & Filter & $t_{\rm exp}$ & $\mu$-step & Offsets & $t_{\rm tot}$ \\
(Survey) &       & (s)          &         &         &   (s)        \\
 \hline
1 (LAS) & $Y$   & 20 & no         &  2-pt &  40 \\
        & $J$   &  5 & $2\times2$ &  2-pt &  40 \\
        & $HK$  & 10 & no         &  4-pt &  40 \\ \hline
2 (LAS) & $Y$   & 20 & no         &  4-pt &  80 \\
        & $J$   & 10 & $2\times2$ &  2-pt &  80 \\
        & $HK$  & 10 & no         &  8-pt &  80 \\ \hline
3 (GCS) & $ZY$  & 20 & no         &  2-pt &  40 \\
        & $JH$  & 10 & no         &  4-pt &  40 \\
        & $K$   &  5 & $2\times2$ &  2-pt &  40 \\ \hline
4 (GPS) & $JH$  & 10 & $2\times2$ &  2-pt &  80 \\
        & $K$   &  5 & $2\times2$ &  2-pt &  40 \\ 
        & $H_2$ & 20 & $2\times2$ &  2-pt & 160 \\ \hline
5 (DXS) & $JK$  & 10 & $2\times2$ & 16-pt & 640 \\ \hline
6 (UDS) & $JK$  & 10 & $3\times3$ &  9-pt & 810 \\ \hline
\end{tabular}
\caption{The implementation schemes used for the five surveys in
  05B.}
\label{tab_obs_design}
\end{table}

\subsection{Implementation}

The basic observational unit of the surveys is the stack multiframe,
which is the set of four frames (one for each detector) formed by
combining the set of exposures, of individual length $t_{\rm exp}$, and
combined length $t_{\rm tot}$, at a given base position. The shallow
surveys, LAS, GCS, and GPS, are built up by tiling the survey areas with
these stack frames. For the deep surveys, DXS, and UDS, deep stack frames
are created by averaging several stack frames at each position, and
the fields are tiled with these deep stack frames. 

The observing strategy incorporates flexibility, in order to make the
best use of the observing time allocated. For example, projects
suitable for execution in non--photometric conditions are included, as
well as when the seeing is mediocre in photometric conditions. Small
changes were made to the observing strategy for the 05B block in the
light of experience in the 05A block for the LAS, GCS, and DXS. The
revised set of implementation schemes for 05B (update of Table 4 in
D06) is provided in Table \ref{tab_obs_design}. The same scheme has
been retained for subsequent observations, to date. For each scheme,
the table lists successively a reference number, the filters to which
the details apply, the exposure time, the
microstepping\footnote{$2\times2$ or $3\times3$ interlacing of frames
  taken with $n+1/2$ or $n+1/3$ pixel offsets \---\ see D06 for
  details} (if used), the number of offset positions (other than
microsteps), and the total integration time $t_{\rm tot}$ making up
the stack frame.

The changes since 05A are as follows. For the DXS exposure times were
increased from 5s to 10s, to reduce overheads associated with
readout. For the GPS the exposure times in $J$ and $H$ were increased
from 5s to 10s, to reduce the overheads, and the 4-pt offset pattern
replaced by a 2-pt pattern. Observations with the $H_2$ filter, used
for the first time in 05B, mimic the $JH$ procedure, but with twice
the exposure time, 20s, giving a total integration time of 160s. For
the LAS and GCS, again in order to reduce overheads, by allowing
increased exposure time, microstepping is now only applied in a single
filter, $J$ for LAS, and $K$ for GCS. These are the filters where
second--epoch observations will be made, for measuring proper
motions. The improved sampling provided by microstepping allows more
accurate astrometry.

\subsection{Pipeline and calibration}

\subsubsection{Pipeline}
\label{pipe}

The main changes to the pipeline processing since the EDR are a 
modification of the sky subtraction strategy and the incorporation
of an algorithm to reduce the impact of cross-talk images.

The changes to the sky subtraction strategy involved grouping the sky
estimation and correction by exposure time within each passband, if
possible.  This was not necessary for the EDR since virtually all
frames in any one passband used identical exposure times.  This
extra refinement proved necessary due to the presence of a combination
of additive and multiplicative artifacts in the sky correction
frames. The generally more localised multiplicative artifacts scale
closely with exposure time whereas the additive components do not. The
latter includes illumination-dependent reset anomaly and pedestal
offsets. Because a wide range of observing strategies may be employed
over a night, the manner in which to combine data over the night to
optimise the sky subtraction is a complex problem. It has proven
difficult to devise a general algorithm that does not require manual
intervention, and in a few percent of images sky-subtraction residuals
are visible. Further improvements to the sky-subtraction algorithm are
anticipated.

Cross-talk artifacts are confined within detector quadrants and occur
between the eight channels read out in parallel in each quadrant.  All
the detectors show similar cross-talk patterns with the induced
artifacts being essentially time derivatives of saturated stars, with
either a `doughnut' appearance from heavily saturated regions or
half-moon-like (i.e. positive/negative residual) appearance from only
weakly saturated stars.  Adjacent cross-talk images are symmetric
(with number of channels displacement) and typically induce features
at $\sim1$ per cent of the differential flux of the source in adjacent
channels, dropping to $\sim0.2$ per cent and $\sim0.05$ per cent in successive
channels further out. The size of the effect continues to decline with distance
from the source, but in the case of very bright stars in the first or
eighth channel, may be detectable out to seven channels distance.

Creating a correction for these artifacts involves devising a robust
combination of a model involving the directional time derivative of
the primary source, and if possible, comparing and combining this with
adjacent cross-talk artifacts to create a correction sub-image.  In
general this model sub-image is unaffected by the presence of real
objects overlapping the artifact and although the time derivative
model is only an approximation to the real effect a relatively clean
subtraction generally results.

D06 noted two problems affecting stacking of the data in the deep
surveys. For the DXS and UDS the basic data product is a {\em
  leavstack} multiframe (i.e. four detectors) of total integration
time 640s and 810s respectively (Table \ref{tab_obs_design}), referred
to here as an intermediate stack.  Depth is built up by averaging
these intermediate stack frames. The first problem was caused by a
safety feature being incorrectly applied, which meant that the
recorded value of the sky noise in intermediate stacks was set too
high in a few cases. The second problem (which did not affect the UDS)
was quantisation noise, due to the use of integer values, affecting
the deepest DXS frames. Revised stacking procedures have eliminated
both problems.

\subsubsection{UDS processing and source detection}

The UDS is the only survey which employs $3\times3$ microstepping. The
resulting pixel scale ($0.133\arcsec$) oversamples the typical seeing
of the UDS images ($0.7\arcsec$). The image detection algorithm
employed in the WFCAM pipeline is optimised for critically sampled, or
slightly oversampled, data. As noted in D06, for highly oversampled
data, such as for the UDS, the algorithm tends to break up objects
into multiple components. Consequently the UDS EDR object catalogues
were unsatisfactory in this respect. Because of this, new pipeline
procedures have been implemented for the UDS for DR1. The standard
WFCAM pipeline is followed, up to creation of the intermediate stack
frames. From that point on, further processing, quality control, frame
stacking, and catalogue generation have been undertaken by the UDS
working group. Full details will be provided in a forthcoming paper
\citep{foucaud06}. A brief outline of the procedures follows.

All intermediate stacks were visually inspected for defects. In a
substantial fraction the sky subtraction was deemed
unsatisfactory. This can be imputed to the manner in which data taken
at different times in the night were combined to create sky correction
frames (Section \ref{pipe}). The unsatisfactory data were passed
through the pipeline again, manually tailoring the sky subtraction
process, and the majority of the data were recovered. A further
approximately 5 per cent of frames were discarded due to the presence of
moon ghosts (see Section \ref{moon} below). Channel bias offsets (D06)
were identified in a few per cent of the frames and were corrected for
with a custom procedure.  In total 76 per cent of the UDS intermediate stacks
were used in the DR1 release.

Before the final stacking, noisy border regions were trimmed from each
frame using a masking procedure. Masking was also applied to minor
image defects (e.g. satellite trails). The stacking was performed
using the Terapix SWARP image resampling tool \citep{bertin02}, with
flux scaling computed from the calibration zero-points. Images were
weighted in the final stack by the inverse variance of the sky.
Experimentation confirmed that this improved the final depths by
approximately 0.1 mag. The pipeline confidence maps were then used to
provide inter-pixel weighting.

Source catalogues were produced using the SExtractor software
\citep{sextractor}. The {\em K-}band image was used as the primary
detection image, since this is deeper than {\em J} for most galaxy
colours. Simulations were used to optimise the source extraction
parameters, as described in detail in \citet{foucaud06}.

\subsubsection{Calibration}
\label{calib}

The calibration scheme applied for DR1 is exactly the same as used for
the EDR (D06): the zero point for each detector in each stack
multiframe is determined by identifying suitable 2MASS stars in the
frame, and converting the 2MASS magnitudes to the WFCAM {\em ZYJHK}
system using linear colour equations, listed in D06. The calibration
goal is an accuracy of the zero point of 0.02mag. While there are
sufficient 2MASS stars on each detector to achieve a precision better
than 0.02mag, systematic errors can enter in a variety of ways. We
anticipate three changes to the way in which the calibration is
undertaken for DR2.
\begin{itemize}
\item For stars cooler than spectral type K, typically the colour
  relations become non-linear, and differences between dwarfs and
  giants become significant. Therefore in DR2 colour equations will be
  determined, and applied, over a limited colour range, confined to
  hotter stars, F to K.
\item The colour terms for the {\em Z} and {\em Y} bands are large,
  and it has become apparent that these relations are non linear, with
  a break near the boundary between A and F spectral types. Therefore
  a colour relation determined from a linear fit to the F to K sequence will
  result in a constant zero point error. Preliminary analysis
  indicates that the offset required is larger in the {\em Y} band
  than in the {\em Z} band, and is approximately 0.09mag, in the sense
  that {\em Y} magnitudes in the EDR and DR1 should be increased by
  this amount to place them on the Vega system. Following a more
  detailed analysis, the DR2 photometry will be corrected for this
  source of error.
\item There is evidence of zero-point errors in regions of high
  extinction. This is simply a consequence of the fact that the colour
  relations for extinguished stars will differ from the colour
  relations for unextinguished stars, and will depend on the
  extinction. Therefore in highly extinguished regions application of
  a colour equation determined from regions of low extinction is
  incorrect.
\end{itemize}

\subsection{Data artifacts}
\label{moon}

The cause of the bright moon ghosts described by D06 has been traced
to the autoguider auxiliary lens that sits at the centre and above the
field lens \citep{casali06}. When the moon lies within an annulus from
about $14^{\circ}$ to $30^{\circ}$ off axis, a ghost is formed by
moonlight shining through the auxiliary lens and undergoing a double
internal reflection inside the field lens.  It is planned to install a
baffle tube between the auxiliary lens and the field lens to eliminate
these moon ghosts. In the meantime, over the 06A observing block,
fields within $30^{\circ}$ of the moon have been avoided.  Because the
arrays cover only a fraction of the focal plane the moon ghosts also
produce scattered light in the instrument, even when no ghost image
occurs. Sky subtraction is frequently poor when the moon is within
$30^{\circ}$ of the field, although it is not clear whether this is a
general problem caused by the moon shining directly on the field lens,
or whether it is a consequence of the ghost images, and scattered
light within the instrument. Whatever the cause, the majority of
images taken in 05A and 05B, when the moon was within $30^{\circ}$,
have been deprecated.

\subsection{Quality control}

The QC procedures followed the same pattern as with the EDR, {\em vis.}
first removing corrupt data (i.e. meaningless data, e.g. empty array),
and bad data (i.e. unusable data, due e.g. to a moon ghost), and then
applying a set of QC cuts that define survey quality. The last have
now been finalised and we set out the main elements here. The most
important parameters defining survey quality are photometric zero
point (i.e. how much cloud), average stellar image ellipticity,
seeing, and depth. The criteria are detailed below. Some summary
statistics quantifying the characteristics of the data are provided in
Section \ref{summary}.

{\bf Photometric zero point.} Each detector frame is calibrated using
the 2MASS photometry of bright unsaturated stars in the frame, using
appropriate colour terms. Therefore the computed zero point for each
frame, relative to the modal value for that filter, gives an
indication of how much extinction by cloud there was.  The effect of
cloud is to reduce the depth of the observations, but cloud only
effects the accuracy of the calibration to the extent that the
extinction is variable across the field of view. Therefore, perfectly
photometric conditions are not necessarily required for good
calibration.

In practice all projects requiring accurate calibration specify
`photometric' conditions, and extinction by cloud is monitored during
execution. Inevitably, though, a fraction of these observations are
undertaken through thin cloud.  At the QC stage, for photometric
projects we require that the zero point lie within 0.2\,mag. of the
modal value. In practice the vast majority of observations requesting
photometric conditions are undertaken in conditions considerably
better than this limit: the root mean square (RMS) variation of the
zeropoint for frames specifying photometric conditions is 0.02\,mag in
{\em JHK}.

For some projects photometric conditions are not required, and these
are specified as `thin cirrus' at the telescope. These include DXS and
UDS observations, since each field is visited repeatedly, and the
field may be calibrated if only a single visit takes place in
photometric conditions. Additionally `repeat' observations in the
shallow surveys ({\em J} only in the LAS, {\em K} only in the GCS, and
{\em K} and $H_2$ in the GPS) may be undertaken in non--photometric
conditions.\footnote{These observations are scheduled to take place at
an interval of greater than two years relative to the primary
observations in the full filter set. Sometimes they are scheduled
before and sometimes after.}  At the QC stage, for non-photometric
projects we require that the zero point lie within 0.3\,mag. of the
modal value.

\begin{figure}
\includegraphics[width=8cm]{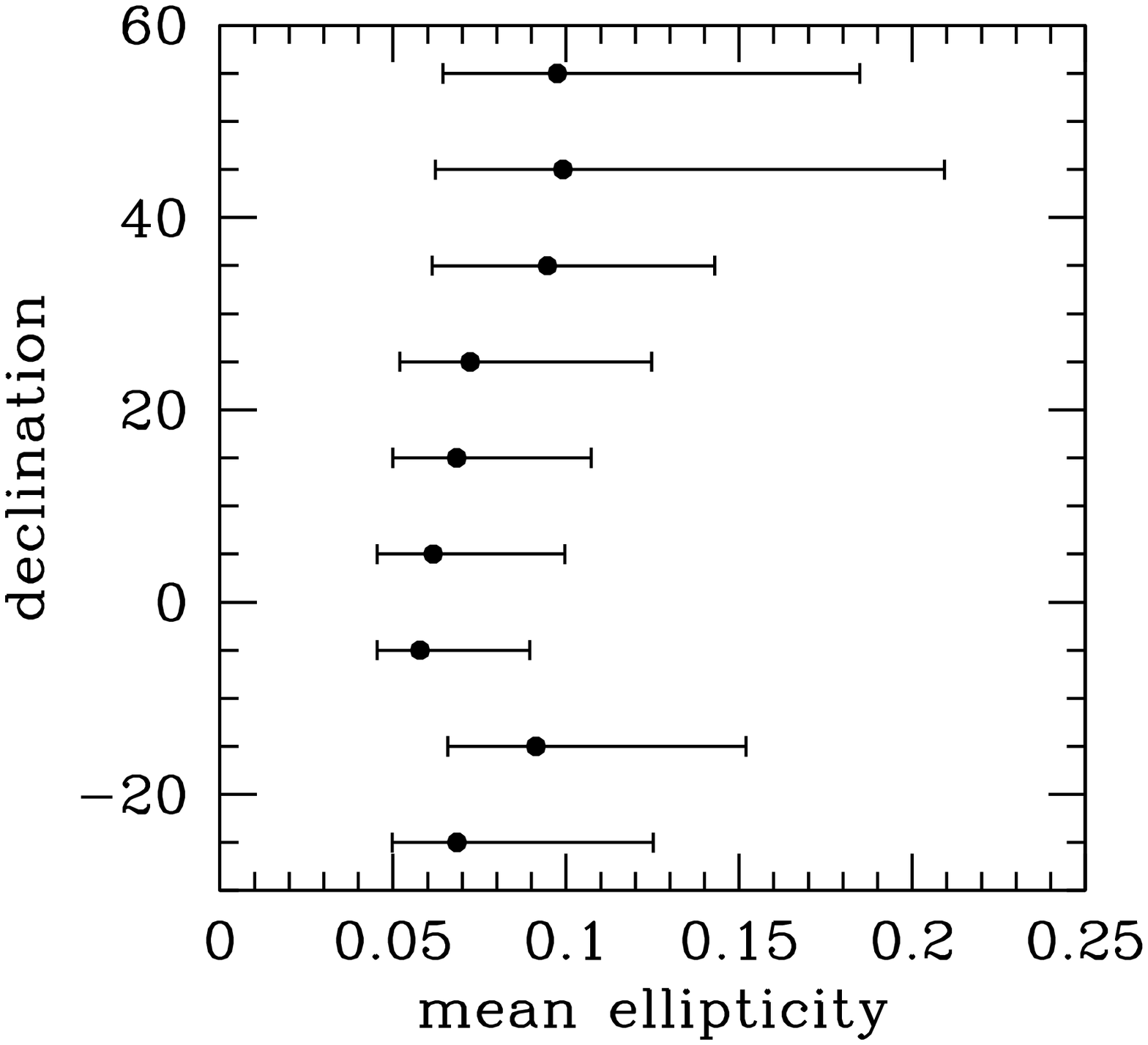}
\caption{Plot of distribution of stellar ellipticity, averaged over
  each multiframe, within declination slices. Each point marks the
  median value in all multiframes within the slice, and the error
  bars denote the 16 per cent to 84 per cent range. Some frames at high
  declinations suffer from elongated images.}
\label{ellipplot}
\end{figure}

{\bf Ellipticity.} The mean value of the ellipticity $(e=1-b/a)$ for all
the stellar objects in a detector frame is recorded in the archive as
the parameter {\tt avStellarEll} in the table {\tt
  MultiframeDetector}. The value of this parameter averaged over the
four detectors of a multiframe is used as a QC parameter. In the
majority of cases the measured value is less than 0.1, but a
proportion of the DR1 frames display elongated images, particularly at
high declinations. The extent of the problem is illustrated in
Fig. \ref{ellipplot} which plots the distribution of the ellipticity
parameter over all multiframes observed for DR1 (before any QC
deprecations), within slices of declination. The points are the median
value, and the error bars mark the 16 per cent to 84 per cent range (equivalent
to $\pm1\sigma$). At high declinations $\delta>40^{\circ}$ there is a
pronounced tail towards large values. For projects at high
declinations we deprecated multiframes with stellar ellipticity
$>0.25$, while a cut $>0.20$ was applied to all other frames.

The cause of the elliptical stellar images has been traced to
instrument flexure, as well as uncorrected errors in the primary
mirror figure, introducing astigmatism and coma. These problems of
image quality were solved in time for 06A observations. Using a model
derived from wave-front sensing data, the effects of instrument
flexure are now corrected for by tilting and displacing the
secondary. Additionally, the telescope temperature and
attitude--dependent focus model has been improved for the combined
UKIRT/WFCAM system leading to better focus tracking during observing.

{\bf Seeing.} For the LAS and GCS we required seeing $<1.2\arcsec$
(averaged over the four detectors), except for the single-band
non-photometric observations ({\em J} in LAS, {\em K} in GCS) where
the seeing was relaxed to $<1.3\arcsec$. For the GPS, in the plane
(i.e. except for observations of Taurus), the requirements were
$<1.0\arcsec$ in {\em K}, and $<1.1\arcsec$ in {\em J} and {\em H}. In
Taurus the limit was relaxed to $<1.3\arcsec$ in {\em K} and {\em
  H$_2$}, and $<1.4\arcsec$ in {\em J} and {\em H}. In the DXS the
limit was $<1.3\arcsec$, and in the UDS the limit was $<0.8\arcsec$ in
{\em K}, and $<0.9\arcsec$ in {\em J}.

{\bf Depth.} For the LAS and GCS, detector frames are deprecated if
the depth achieved (as defined by D06) is more than
0.5\,mag. less deep than the modal value. For the GPS, the depth is
affected by confusion noise, and therefore depends on Galactic
latitude, $b$. So, for the GPS, depth is plotted against $b$, and
outliers are identified and deprecated. There are no specific depth
cuts applied to the DXS and UDS observations, since all frames
contribute to improving the depth in the deep stack
frames. Nevertheless a small number of frames are eliminated that were
taken in conditions of bright sky, where the depth is severely
compromised.

{\bf Visual checks.} An additional check is made of each stack frame
by viewing the archive jpeg image to identify bad frames not already
deprecated by the automated QC procedures. Examples include frames
where the sky subtraction is visually unsatisfactory, trailed frames
not already identified as such, frames suffering from extreme channel
bias offsets (either individual channels or multiple channels), or
frames badly affected by bright moon ghosts. Where only a small part
of a moon ghost affects the frame, or where the ghost is not
particularly bright, the frame may be accepted.

{\bf Proportion of frames failing quality control.} Summed over all
the surveys the proportion of frames that failed QC is close to
$20\%$. We expect this figure to fall substantially in future
releases, to $<10\%$. The main causes are the following: {\em a.)}
$7\%$ of frames were duplicated, because the observations were
abandoned (principally because the seeing deteriorated), and were
later repeated; {\em b.)} $8\%$ of frames were rejected because the
sky subtraction was unsatisfactory, or bright moon ghosts were
present; {\em c.)} $4\%$ of frames were rejected because the seeing
exceeded the required limits, or the images were elongated.

We expect the proportion in category {\em (a)} to be substantially
smaller in future releases. If the seeing deteriorates during an
observation the observer may abandon the MSB. Unfortunately an earlier
version of the summit pipeline failed to keep pace with the
observations, when the GPS was observed (because cataloguing the
myriad stars was slow), meaning that the QC information was only
available after completion of the MSB, so the whole MSB was rejected,
when it should have been abandoned earlier. The summit pipeline now
keeps up with the observations. Similarly the proportion in category
{\em (b)} will be smaller, now that the moon ghosts have been
eliminated, and with improvements to the sky subtraction
algorithm. There should also be some reduction in category {\em (c)},
with image-quality improvements. Nevertheless failure of a small
fraction in this category is considered acceptable, recognising this
as a consequence of attempting to use as large a fraction of the
observing time as possible, by tailoring the observations to the
conditions.

\subsection{Archive}
\label{archivesec}

Between the EDR and DR1 releases, there have been several minor bug
fixes to the archiving software. These are detailed in the release
history notes in the online web pages. Some modifications, however,
warrant elaboration.

In the EDR, there were a small number of fields where frames were not
correctly associated, and this gave rise to incompletely merged
sources e.g.  the {\em YJHK} bands available in LAS fields being
merged into two {\em framesets}, for example one comprising {\em YJH}
and the other {\em K} only. This in turn led to a small but
significant number ($10^3$ out of $10^6$, or 0.1 per cent) of merged
sources of apparently extreme colour when trawling the EDR+ tables for
objects appearing in a limited set of the available bands. This bug
has been fixed in the DR1 release. At the same time the source merging
software (Hambly et al., 2007, in prep.), and the algorithm that
associates detector frames into framesets have been refined. As a
consequence, for some sources the choice of detector frames making up
a frameset has changed between EDR and DR1, and therefore so also has
the photometry.

As noted previously, the treatment of both the DXS and UDS in the EDR
was suboptimal, mainly in terms of the source extraction applied to
each.  The code used was the standard nightly pipeline code, and this
was not set up to cope with the lower sky noise of the deep surveys
(DXS, UDS), or the significant oversampling of the UDS data. For the
DXS, a revised procedure of scaling the data numbers before performing
source extraction works around the limitations in the software; for
the UDS, we have adopted the data yielded by the alternative software
and procedures detailed in \citet{foucaud06}. No changes to the
standard archive data model were required by either of these
procedural modifications, but the online documentation for the UDS has
been substantially altered to correctly describe the changes to the
archive web pages.

\section{Fields observed in DR1}
\label{dr1}

In this section we provide maps of the areas covered in DR1. In cases
where the maps provided in D06 have not required updating, they are
not repeated here. This is made clear when describing each survey.

\begin{figure*}
\caption{Coverage by LAS 05B observations, showing data coverage in
  the DR1 database (all of {\em YJHK}; dark--grey tiles only), and the
  DR1+ database (all tiles). Each small square is a detector
  frame. The physical size of a detector in this plot and in the GCS
  plot (Fig. \ref{gcs05b}), is twice as large as in the two GPS plots
  (Figs \ref{gpseast} and \ref{gpswest}).}
\label{las05b}
\end{figure*}

\subsection{Large Area Survey}

In 05A the projects observed were LAS1, 2, 3, 4. The maps for these
projects are provided in D06. New fields were observed in 05B and are
illustrated below. Together with the EDR fields they form DR1.

The new fields lie within projects LAS$5-11$, which are illustrated in
Fig. \ref{ukidss_survey_areas_dr1}. Projects LAS$5-8$ cover successive
RA sections of the SDSS southern equatorial stripe 82. Additional
early--spring projects LAS$9-11$ were observed towards the end of
05B. Projects LAS$5-9$ specified photometric, and good conditions (seeing,
sky brightness), while project LAS10 specified photometric, and poor
conditions, and integration times were doubled. Project LAS11 was
specified for thin cirrus conditions, with double integration times,
and was observed in {\em J} only.

The coverage achieved in each project is illustrated in
Fig. \ref{las05b}. In the plots every small square represents a detector.
A dark-grey square denotes data in the DR1 database, i.e.  where the
full filter complement {\em YJHK} is complete. A light-grey square
denotes additional data in the DR1+ database, where the filter
complement is incomplete. Observations are undertaken in Minimum
Schedulable Blocks (MSBs, see D06) lasting about an hour. The LAS MSBs
typically consist of coverage of five tiles in a pair of filters,
either {\em YJ} or {\em HK}. In 05B coverage was much more complete in
the {\em HK} pairs. This is a consequence of the restriction on sky
brightness imposed on the {\em YJ} observations.

A problem in locating bright enough guide stars resulted in a quirk
affecting a few frames in projects LAS5 and LAS7. In the archive, the
source merging algorithm associates detector frames in different
filters that are spatially coincident, forming a frameset, and
then matches detections over the filter set. Some {\em YJ}
observations, which tile the same area, were made on different centres
to the {\em HK} observations. The result of the source merging
procedures is that there are a number of sources recorded in the
archive as detected in {\em YJ} only, coincident with distinct sources
(in reality the same sources) recorded as detected in {\em HK}
only. It is intended to rationalise such occurrences in future
releases.

\begin{figure*}
\caption{Coverage by GCS 05B observations, showing data coverage in
  the DR1 database (all of {\em ZYJHK}; dark--grey tiles only), and the
  DR1+ database (all tiles). Each small square is a detector
  frame. Because $\alpha$ and $\delta$ are plotted as rectangular
  coordinates, the detectors appear distorted by sec$\delta$ at high
  declinations. The physical size of a detector in this plot and in the LAS
  plot (Fig. \ref{las05b}), is twice as large as in the two GPS plots
  (Figs \ref{gpseast} and \ref{gpswest}).}
\label{gcs05b}
\end{figure*}

\subsection{Galactic Clusters Survey}

The GCS will observe the 10 star clusters listed in Table 3 in D06.
In 05A the targets Sco and Coma--Ber were observed. The maps for these
projects are provided in D06. New fields were observed in 05B and are
illustrated below. Together with the EDR fields they form DR1.

The five new targets observed in 05B are the Pleiades, Alpha--Per,
Tau.--Aur., Orion, and the Hyades. The locations of the targets are
illustrated in Fig. \ref{ukidss_survey_areas_dr1}, and coverage maps are
provided in Fig. \ref{gcs05b}. The Hyades observations were in the
{\em K} band only. For the other four targets, dark--grey squares
denote data in the DR1 database, where coverage in the full filter
complement {\em ZYJHK} exists, and light-grey squares denote
additional data in the DR1+ database, where the filter coverage is
incomplete.

\begin{figure*}
\caption{Coverage plot for the GPS eastern wing, showing data coverage
  in the DR1 database (all of {\em JHK}; dark--grey tiles only), and
  the DR1+ database (all tiles). Each small square is a detector
  frame. Because $\alpha$ and $\delta$ are plotted as rectangular
  coordinates, the detectors appear distorted by sec$\delta$ at high
  declinations. The light shaded tiles at $|b|>1$ were deliberately
  observed at {\em K} band only, in non-photometric conditions.  {\em
    JHK} observations were attempted only at $|b|<1$, in photometric
  conditions. The physical size of a detector in the two GPS plots is
  half as large as in the LAS (Fig. \ref{las05b}) and GCS plots
  (Fig. \ref{gcs05b}).}
\label{gpseast}
\end{figure*}

\begin{figure*}
\caption{Coverage plot for the GPS western wing, showing data coverage
  in the DR1 database (all of {\em JHK}; dark--grey tiles only), and
  the DR1+ database (all tiles). Each small square is a detector
  frame. Because $\alpha$ and $\delta$ are plotted as rectangular
  coordinates, the detectors appear distorted by sec$\delta$ at high
  declinations. The irregularly shaped outline of the
  Taurus-Auriga-Perseus complex follows the outer contour of the CO
  gas map of \citet{tapregion}, but avoids the $b=-5$
  limit of the main survey. The physical size of a detector in the two
  GPS plots is half as large as in the LAS (Fig. \ref{las05b}) and GCS
  plots (Fig. \ref{gcs05b}).}
\label{gpswest}
\end{figure*}

\subsection{Galactic Plane Survey}

The coverage plots for the eastern and western wings of the GPS are
provided in Figs \ref{gpseast} and \ref{gpswest}. The $H_2$
observations are confined to the Taurus--Auriga--Perseus field.
For the GPS the DR1 database is defined by fields covered by the three
filters {\em JHK} i.e. without regard to $H_2$ coverage.

The fields at Galactic latitudes $1<|b|<5$ shown in Fig. \ref{gpseast}
were only observed in the $K$ band, in non-photometric conditions, and
therefore have light shading. Fields at $|b|<1$ were observed in the
{\em JHK} bands in photometric conditions and have dark
shading, provided that data passed quality control on all three
bands. The regions at $1<|b|<5$ will be observed at {\em JHK} later in
photometric conditions, in line with the GPS plan to observe all
regions at three separate epochs, twice in just the $K$ band, and once
at {\em JHK}. It is clear from Fig. \ref{gpseast} that relatively
little {\em JHK} data were taken in the ``inner Galaxy'' region at
$-2<l<107$, $|b|<1$, largely due to poor weather. However this
situation is expected to improve dramatically in the next UKIDSS
release, DR2, following a recent successful observing season in summer
2006.

Fig. \ref{gpswest} shows the observations in the ``outer Galaxy'' region at
$141<l<230$, as well as the coverage of the Taurus-Auriga-Perseus
complex, where $H_2$ observations are included.  In the outer Galaxy
survey design there is no differentiation by Galactic latitude. Some
areas of the outer Galaxy region (defined by declination) were
selected for {\em JHK} photometric observation. These are shown as
dark tiles if data in all three filters passed quality control.  Other
areas were selected for only {\em K} band observation, in poor weather
conditions, and are shown as light tiles. The observations of the
Taurus-Auriga-Perseus complex were similarly split into photometric
({\em JHK+H}$_2$) and non-photometric ({\em K+H}$_2$)
regions. Observations in the photometric section in the southernmost
part of the complex are shown as dark tiles if data in all of {\em JHK}
passed quality control. Observations in the other, non-photometric
sections, are shown as light shaded tiles.  Again, it is intended to
observe all regions in Fig. \ref{gpswest} with the full complement of {\em JHK}
or {\em JHK+H}$_2$ filters eventually.

\begin{figure*}
\caption{DXS coverage and depth in {\em J}. Tiles that are available
  from the WSA as {\em deepleavstacks} are labelled. Grey-shaded key
  gives depth in {\em J}.}
\label{dxsj}
\end{figure*}

\begin{figure*}
\caption{DXS coverage and depth in {\em K}. Tiles that are available
  from the WSA as {\em deepleavstacks} are labelled. Grey-shaded key
  gives depth in {\em K}.}
\label{dxsk}
\end{figure*}

\subsection{Deep ExtraGalactic Survey}

The DXS covers four fields (Section \ref{projects}) in {\em J} and
{\em K}. The coverage plot for {\em J} is provided in Fig. \ref{dxsj},
and for {\em K} in Fig. \ref{dxsk}. In these plots depth is
illustrated by tone, where darker means deeper. The basic data product
for the DXS is a {\em leavstack} multiframe (i.e. four detectors) of
total integration time 640s (Table \ref{tab_obs_design}).  Depth is
built up over several visits, by averaging these intermediate stack
frames. The aim is to reach full depth \citep[$J=22.3$,
  $K=20.8$,][]{lawrence06} over a full tile, before moving to the next
tile. Once several intermediate stacks exist, they are combined in the
archive to form {\em deepleavstacks}, one for each detector. The tiles
for which {\em deepleavstacks} have been created are marked as `deep'
in Figs \ref{dxsj} and \ref{dxsk}. Details of the depths reached are
provided in Section \ref{summary}. Merging of sources across bands in
the DXS has only been undertaken for {\em deepleavstack} frames.

\subsection{Ultra Deep Survey}

The UDS consists of a single tile, 0.8deg$^2$, centred on $2^{\rm h}18^{\rm m},$
$-5^{\circ}10^\prime$. The observing strategy builds up depth
uniformly over the tile. The UDS was observed only in 05B, in the {\em
  J} and {\em K} bands. Because the depth coverage over the tile is
consequently quite uniform, we have not provided a coverage
map. Details of the depths reached are provided in Section
\ref{summary}.

\section{Summary of the contents of DR1}
\label{summary}

\begin{figure}
\includegraphics[width=8.0cm]{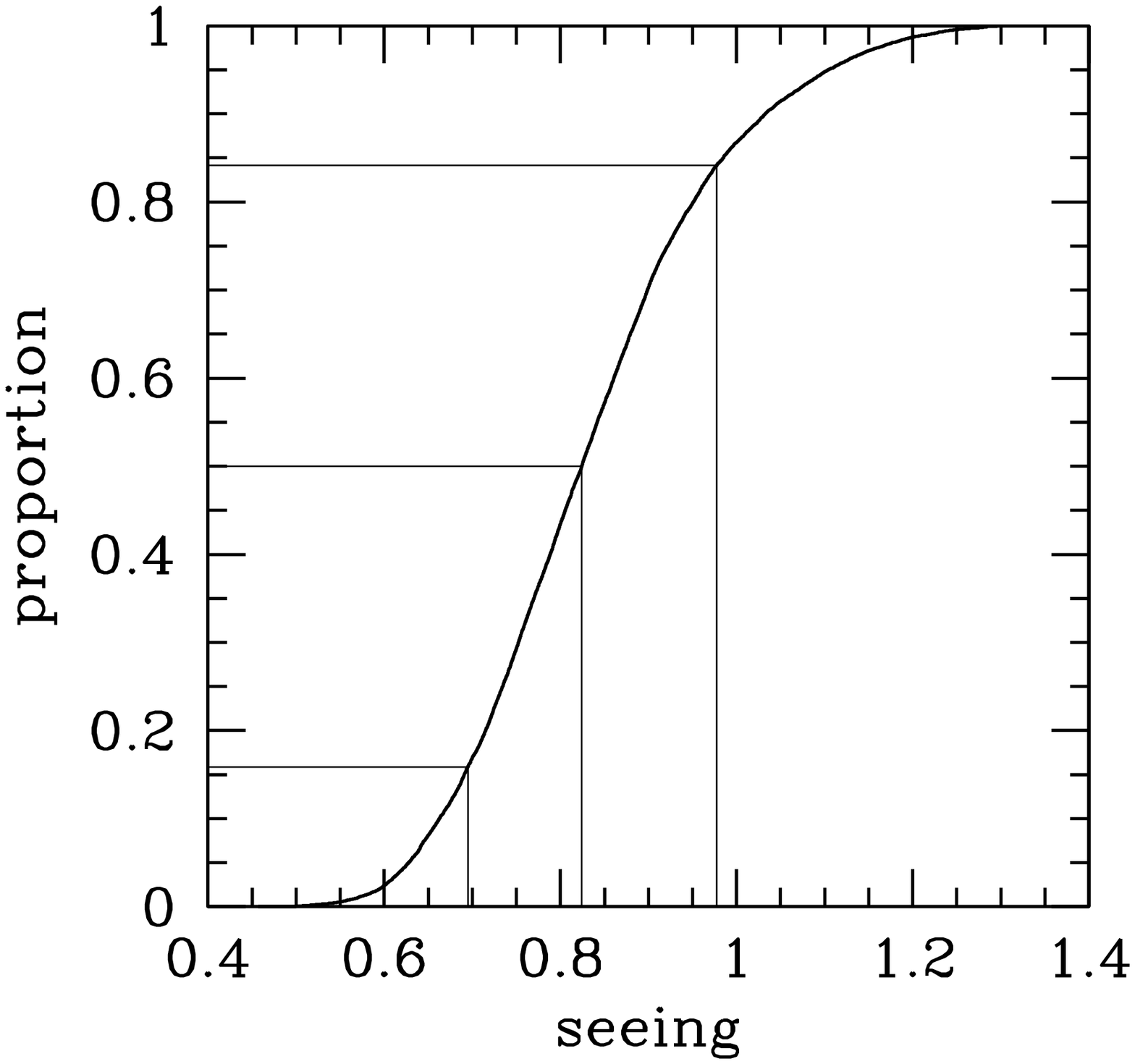}
\caption{Plot of the cumulative distribution of the seeing measured in
all the detector frames in the DR1+ database.}
\label{seeing}
\end{figure}

In this section we provide a summary of the contents of DR1. First we
summarise some general characteristics of the data, followed by
details of the shallow surveys, then details of the deep surveys, and
finally we quantify the overall progress of the surveys towards the
goals set out in \citet{lawrence06}.

\subsection{General details}

An indication of the quality of the data is provided by the cumulative
seeing distribution for all the observations in DR1, reproduced in
Fig. \ref{seeing}. The median value, and $1\sigma$ range (i.e. the 16
and 84 per cent quantiles), is $0.82^{+0.15}_{-0.13}$. The corresponding
quantity for stellar ellipticity is $0.07^{+0.04}_{-0.02}$. Some small
improvements can be expected in both quantities in the future, since
the 05A data suffered from imperfect alignment of the optics, and
because compensation for instrument flexure was not implemented until
after completion of the 05B observations.

\begin{table}
\centering
\begin{tabular}{@{}ccc@{}}
\hline
Colour & RMS spread & no. detectors \\ \hline
$Y-J$ & 0.022 & 4632 \\
$J-H$ & 0.021 & 4394 \\
$H-K$ & 0.026 & 6411 \\
\hline 
\end{tabular}
\caption{Uniformity of photometric calibration, quantified by the
  spread of colours from field to field in the LAS. The quantity
  tabulated for each colour is the standard deviation of the median
  colour of stars in each detector, followed by the number of
  detectors used for the calculation.} 
\label{colourshifts}
\end{table}

A measure of the integrity of the photometry is provided by examining
the change in the average colour of stars from field to field. To
avoid fields with large extinction, we confined this analysis to the
LAS. Then, for colours $Y-J$, $J-H$, $H-K$, for each detector, we
extracted from the database the colours of all stars within a broad
colour range ($0.0-1.0$ for $Y-J$, $0.0-0.7$ for $J-H$, $H-K$), with
colour errors $<0.1$, and recorded the median value. We then computed
the RMS value for the distribution over detectors. The results are
summarised in Table \ref{colourshifts}. The RMS shifts are in the
range $0.02-0.03$\,mag. There is likely to be a contribution to the
spread of colours from the change in the population mix with Galactic
coordinates, and a stochastic contribution due to the limited number
of stars on each detector. This analysis indicates that the uniformity
of the calibration of colours in the survey is better than 0.03\,mag.,
implying that the uniformity of the calibration in a single band is
better than 0.02\,mag. This demonstrates the high degree of uniformity
of the 2MASS calibration \citep{nikolaev00}, and that our procedures
for calibrating from 2MASS are themselves not introducing significant
errors. In D06 we obtained an estimate of the uniformity of the
calibration in a single band of 0.04mag, established by comparing
photometry in the overlapping regions between frames. But the
calibration might be expected to be worst at the detector edges, so we
consider the colour-based measure to be more representative.  The
foregoing analysis, nevertheless, does not preclude the possibility of
a uniform zero-point error across the survey in any band (see Section
\ref{calib}). Neither does it preclude the possibility of systematic
errors being introduced in the calibration of the GCS and GPS, where
reddening, and a different mix of stellar populations relative to the
LAS could require different colour equations.

\begin{figure}
\includegraphics[width=9.0cm]{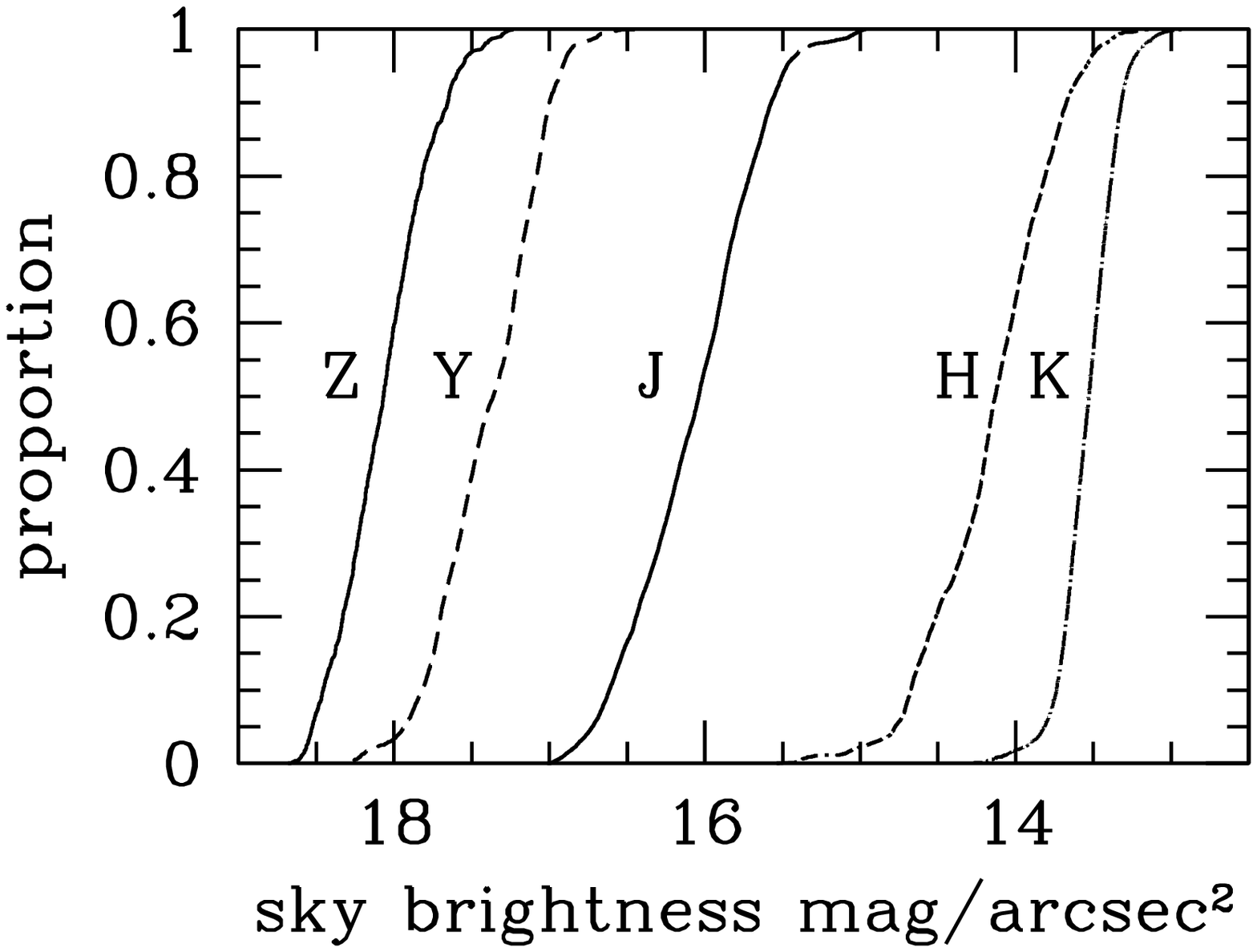}
\caption{Cumulative distribution of the sky brightness in the
  observations in the DR1+ database, by filter {\em Z, Y, J, H, K}.}
\label{skybright}
\end{figure}

\begin{figure}
\includegraphics[width=8.0cm]{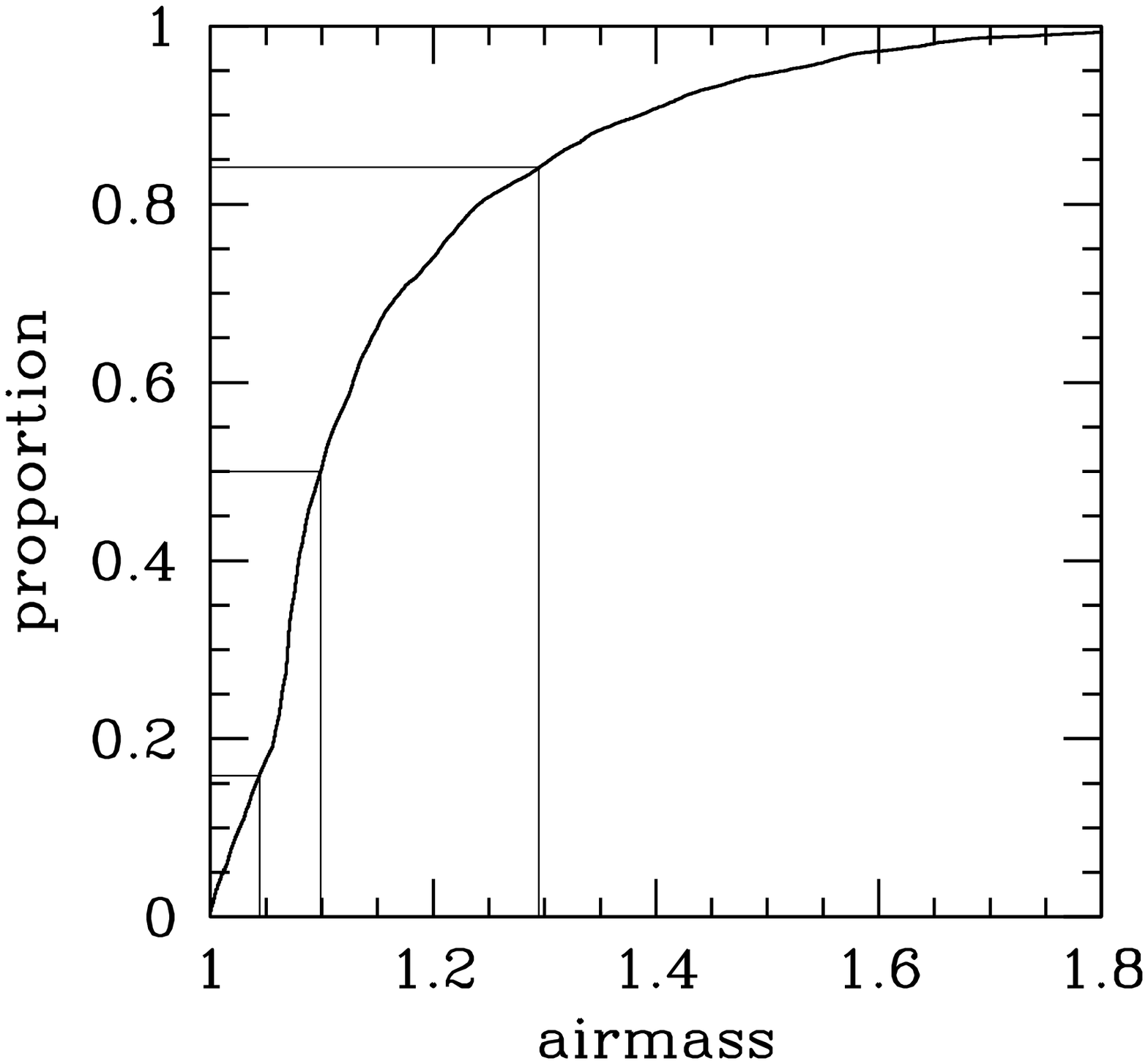}
\caption{Plot of the cumulative distribution of the airmass of all the
  observations in the DR1+ database.}
\label{airmass}
\end{figure}

A plot of the cumulative distribution of the sky brightness in the
observations in DR1 is provided in Fig. \ref{skybright}. The median
values and $1\sigma$ range in each filter are as follows:
$Z_s=18.07^{+0.31}_{-0.31}$, $Y_s=17.36^{+0.38}_{-0.31}$,
$J_s=16.04^{+0.47}_{-0.39}$, $H_s=14.13^{+0.45}_{-0.37}$,
$K_s=13.53^{+0.17}_{-0.17}$mag/arcsec$^2$. Note the smaller range in
{\em K}, because in this filter the background is dominated by slowly
varying thermal emission, rather than the rapidly varying OH emission
which dominates in the other filters. These distributions should be
representative of the night--sky brightness at Mauna Kea, over 2005,
between astronomical twilight, with the exception that the tails to
bright values in the {\em Z, Y, J} bands have been trimmed.  At UKIRT
it has been observed that the sky brightness in the {\em J} band
typically falls rapidly over the first two hours of the
night. Thereafter it follows a similar trend to the {\em H} and {\em
  K} bands, showing, on average, a gentle decline by a few tenths of a
magnitude over the night.  The sky brightness in the {\em J} band is
monitored throughout the night, and observations in the {\em Z, Y, J}
bands specify a maximum sky brightness for execution (D06).

It is noteworthy that the median sky brightness in the {\em Z} band is
0.4\,mag. brighter than the value quoted for SDSS DR1
\citep{abazajian03}. Solar activity, as measured by 2800 MHz flux
\citep[e.g.][]{patat}, does not explain this difference.

It is also of interest to plot the cumulative distribution of airmass
of the observations, which is provided in Fig. \ref{airmass}. Here the
median value, and $1\sigma$ range, is $1.10^{+0.19}_{-0.06}$. This
shows that the scheduling procedures successfully ensure that the
majority of the observations are undertaken at airmasses less than
1.3. Inevitably there is a tail of observations made at higher masses,
because of the declinations of some of the targets.

\subsection{Shallow surveys}

\begin{table*}
\centering
\begin{tabular}{@{}ccccccccc@{}}
\hline
Survey & DR1  & \multicolumn{7}{c}{DR1+ DB} \\
       &  DB  &  $Z$ &  $Y$ &  $J$ &  $H$ &  $K$ & $H_2$ & any \\
\hline
LAS    & 189.6 &   -  & 240.3 & 338.6 & 338.6 & 333.6 &   -   & 474.7 \\
GPS    &  77.2 &   -  &   -   &  88.0 &  88.2 & 343.5 &  70.9 & 361.5 \\
GCS    &  51.8 & 70.9 &  75.9 &  77.6 & 110.1 & 382.2 &   -   & 401.3 \\ \hline
sum    & 318.6 & 70.9 & 316.2 & 504.2 & 536.9 &1059.3 &  70.9 &1237.5 \\
\hline 
\end{tabular}
\caption{Coverage of the shallow surveys (deg$^2$) in the DR1 and
    DR1+ databases.} 
\label{tab_shallow_coverage}
\end{table*}

\begin{table}
\centering
\begin{tabular}{@{}ccccccc@{}}
\hline
Filter & LAS & GPS & GCS \\ \hline
$Z$ &   -   &   -   & 20.36 \\
$Y$ & 20.16 &   -   & 20.05 \\
$J$ & 19.56 & 19.81 & 19.59 \\
$H$ & 18.81 & 19.01 & 18.84 \\
$K$ & 18.19 & 18.07 & 18.16 \\
\hline
\end{tabular}
\caption{The median $5\sigma$ point source depth by filter, in the DR1
  database, for the three shallow surveys, LAS, GPS, GCS.}
\label{tab_shallow_depth}
\end{table}

Details of the summed area covered, by filter, in each of the three
shallow surveys, LAS, GPS, GCS, are provided in Table
\ref{tab_shallow_coverage}. For each survey, the table provides the
area covered in a particular filter, the area covered by all filters
(i.e. the contents of the DR1 database), and the area covered by any
filter. The final row sums up these quantities over the three
surveys. 

Table \ref{tab_shallow_depth} provides the median $5\sigma$ depth (as
defined in D06) achieved in each band in each of the three
surveys. For the LAS and GCS these quantities are similar to those
published in D06 for the EDR. One may anticipate a small increase in
depth in future releases, if, as expected, the image quality
improves. For the GPS these values are on average 0.3\,mag. deeper
compared to the EDR values. All the EDR observations come from the
eastern wing of the GPS, i.e. at low Galactic longitude $l$. The GPS
depths are limited by source confusion, and the effect is most severe
towards the Galactic centre. DR1 includes observations in the western
wing, at larger $l$, where the source number counts are much lower,
resulting in an increase in the median depth in each filter.

\subsection{Deep surveys}

\begin{figure}
\includegraphics[width=7.0cm]{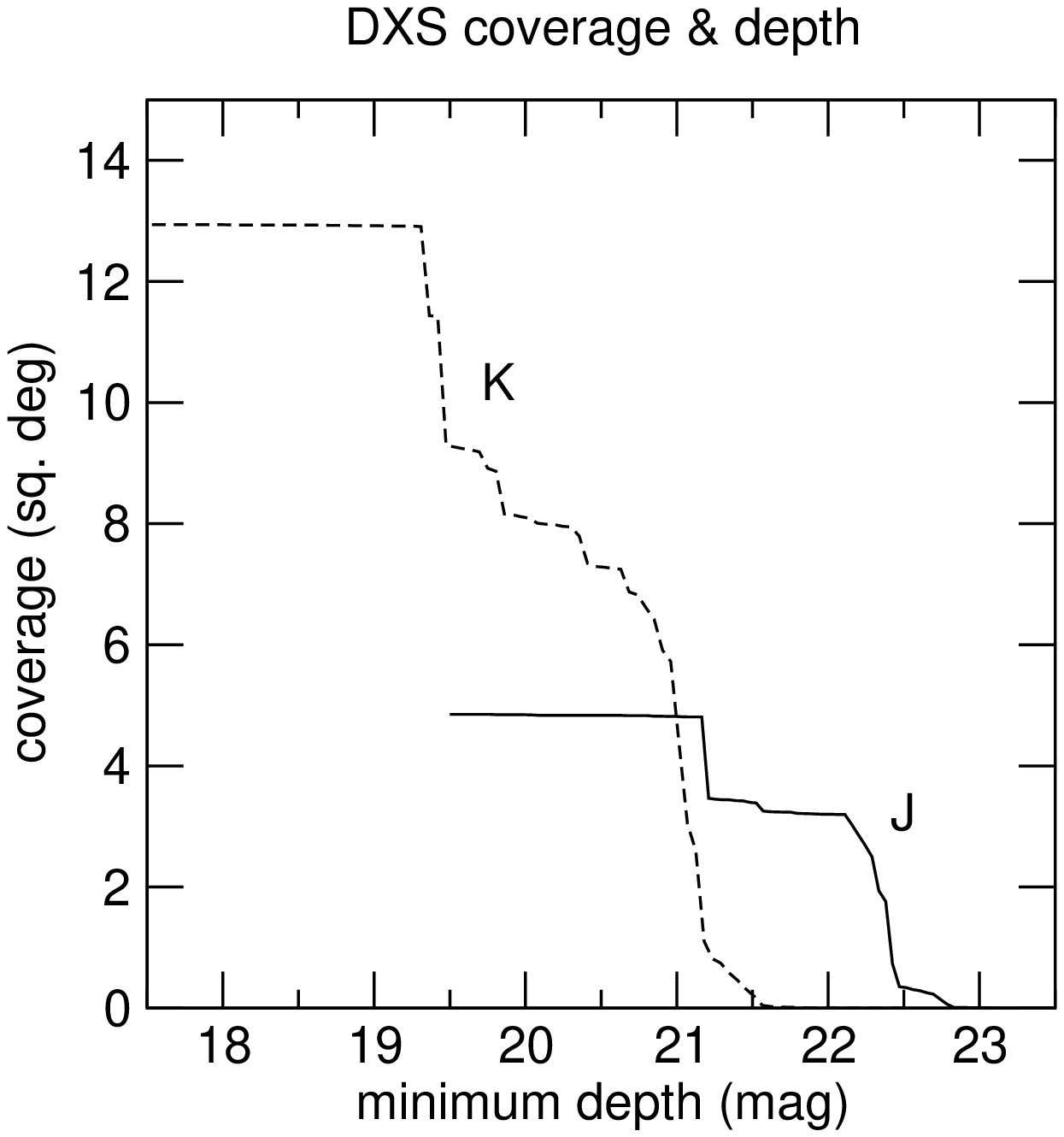}
\caption{DXS areal coverage attaining a given minimum depth as a
  function of the minimum depth, for {\em J} and {\em K}.}
\label{dxsarea}
\end{figure}

\begin{table*}
\centering
\begin{tabular}{|l|c|c|c|c|c|c|c|c|c|} 
\hline
Field &  Sub-field & RA & Dec & $t_{\rm tot}$ (s) &
Depth (mag) & seeing $(\arcsec)$ & $t_{\rm tot}$ (s) & Depth (mag) &
seeing $(\arcsec)$ \\ 
 & & \multicolumn{2}{c}{J2000} & \multicolumn{3}{c}{$J$} & \multicolumn{3}{c}{$K$} \\ 
\hline
XMM-LSS      & 1.00 &  36.5752020 & -4.7496444 &  5760 & 22.26 & 0.88 &  6400 & 20.83 & 0.72 \\
XMM-LSS      & 1.10 &  36.5803542 & -4.5294528 &  6400 & 22.28 & 0.86 &  6400 & 20.87 & 0.74 \\
XMM-LSS      & 1.01 &  36.8012000 & -4.7496444 &  5760 & 22.24 & 0.84 &  6400 & 20.82 & 0.73 \\
XMM-LSS      & 1.11 &  36.8012000 & -4.5294528 &  5760 & 22.27 & 0.85 &  3840 & 20.62 & 0.71 \\
XMM-LSS      & 2.00 &  35.7098417 & -4.7496444 &   -   &   -   &  -   &  5760 & 20.59 & 0.92 \\
XMM-LSS      & 2.10 &  35.7098417 & -4.5294528 &   -   &   -   &  -   &  4480 & 20.54 & 0.86 \\
XMM-LSS      & 2.01 &  35.9306875 & -4.7496444 &   -   &   -   &  -   &  4480 & 20.50 & 0.84 \\
XMM-LSS      & 2.11 &  35.9306875 & -4.5294528 &   -   &   -   &  -   &  5120 & 20.55 & 0.84 \\ \hline
Lockman Hole & 1.00 & 163.3623000 & 57.4753556 &   -   &   -   &  -   & 11460 & 20.94 & 0.93 \\
Lockman Hole & 1.10 & 163.3623000 & 57.6955472 &   -   &   -   &  -   & 10320 & 20.90 & 0.94 \\
Lockman Hole & 1.01 & 163.7756167 & 57.4753556 &   -   &   -   &  -   &  9180 & 20.85 & 0.93 \\
Lockman Hole & 1.11 & 163.7756167 & 57.6955472 &   -   &   -   &  -   &  9180 & 20.86 & 0.95 \\
Lockman Hole & 6.00 & 161.6707620 & 59.1223000 &   -   &   -   &  -   &  4500 & 20.52 & 1.03 \\
Lockman Hole & 6.10 & 161.6707620 & 59.3424917 &   -   &   -   &  -   &  2500 & 20.23 & 0.93 \\
Lockman Hole & 6.01 & 162.1040380 & 59.1223000 &   -   &   -   &  -   &  4000 & 20.48 & 1.01 \\
Lockman Hole & 6.11 & 162.1040380 & 59.3424917 &   -   &   -   &  -   &  3500 & 20.42 & 1.02 \\ \hline
ELAIS N1     & 1.00 & 242.5994000 & 54.5031333 &  2000 & 21.39 & 0.97 &  3000 & 20.25 & 0.93 \\
ELAIS N1     & 1.10 & 242.5994000 & 54.7233250 &  2000 & 21.41 & 0.94 &  4000 & 20.44 & 0.95 \\
ELAIS N1     & 1.01 & 242.9817370 & 54.5031333 &  3640 & 21.64 & 0.96 &  3500 & 20.38 & 0.94 \\
ELAIS N1     & 1.11 & 242.9817370 & 54.7233250 &  2420 & 21.31 & 0.97 &  2500 & 20.17 & 0.95 \\ \hline 
VIMOS 4      & 1.00 & 334.2667460 &  0.1698000 &  7680 & 22.24 & 0.87 &  7340 & 20.74 & 0.81 \\
VIMOS 4      & 1.10 & 334.2667460 &  0.3899917 &  9600 & 22.40 & 0.85 & 11540 & 20.96 & 0.78 \\
VIMOS 4      & 1.01 & 334.4869460 &  0.1698000 &  8960 & 22.38 & 0.82 & 10260 & 20.94 & 0.80 \\
VIMOS 4      & 1.11 & 334.4869460 &  0.3899917 &  5120 & 22.06 & 0.88 &  7480 & 20.75 & 0.83 \\
VIMOS 4      & 2.00 & 335.1420250 &  0.1817444 &  3200 & 21.72 & 0.86 &  7680 & 20.86 & 0.73 \\
VIMOS 4      & 2.10 & 335.1420250 &  0.4019361 &  3840 & 21.80 & 0.89 &  7680 & 20.86 & 0.70 \\
VIMOS 4      & 2.01 & 335.3622250 &  0.1817444 &  2560 & 21.54 & 0.90 &  7040 & 20.80 & 0.70 \\
VIMOS 4      & 2.11 & 335.3622250 &  0.4019361 &  4480 & 21.87 & 0.90 &  7040 & 20.76 & 0.72 \\
VIMOS 4      & 3.00 & 335.1420790 &  1.0559111 &   -   &   -   &  -   &  4480 & 20.42 & 0.87 \\
VIMOS 4      & 3.10 & 335.1420790 &  1.2761028 &   -   &   -   &  -   &  5760 & 20.54 & 0.89 \\
VIMOS 4      & 3.01 & 335.3623330 &  1.0559111 &   -   &   -   &  -   &  4480 & 20.44 & 0.88 \\
VIMOS 4      & 3.11 & 335.3623330 &  1.2761028 &   -   &   -   &  -   &  7040 & 20.68 & 0.87 \\
VIMOS 4      & 4.00 & 334.2668040 &  1.0559111 &   -   &   -   &  -   &  1920 & 19.95 & 0.86 \\
VIMOS 4      & 4.10 & 334.2668040 &  1.2761028 &   -   &   -   &  -   &  3200 & 20.16 & 0.87 \\
VIMOS 4      & 4.01 & 334.4870580 &  1.0559111 &   -   &   -   &  -   &  3200 & 20.26 & 0.92 \\
VIMOS 4      & 4.11 & 334.4870580 &  1.2761028 &   -   &   -   &  -   &  3840 & 20.32 & 0.92 \\
\hline  
\end{tabular}
\caption{The DXS {\em deepleavstack} multiframes. The field, sub-field
  code, and base coordinates, are listed, followed by the total
  integration time in seconds and the 5$\sigma$ point source sensitivity
  in {\em J} and in {\em K}.}
\label{tab_dxs_deep_stacks}
\end{table*}

\subsubsection{Deep ExtraGalactic Survey}

All the intermediate stack frames from 05A and 05B that pass QC have
been included in DR1. However the most important products are the {\em
  deepleavstack} frames. The EDR for the DXS included all the 05A
data, and 05B data taken up to 2005 September 27. Therefore for some of
the deep tiles (e.g. in ELAIS N1) there are no new observations
included in DR1. However the QC and stacking procedures have evolved
since the EDR, so the DR1 stacks supersede any EDR stacks. As already
noted the stacking of {\em deepleavstack} frames for the EDR was less
than optimal.

In DR1 there are nine deep tiles in the {\em K} band, giving 36 {\em
  deepleavstack} multiframes, since there is one such frame per detector.
There are four deep tiles in the {\em J} band, which are
a subset of the {\em K} tiles, and give 16 {\em deepleavstack}
multiframes. Details of all the DXS {\em deepleavstacks} are provided
in Table \ref{tab_dxs_deep_stacks}. The first column provides the
field name, and the second is a code identifying the sub-field, in the
form {\em Tile.XY}, where {\em XY} are binary coordinates specifying one of
four multiframe positions that make up a tile. Columns three and four
provide the coordinates of the base position, which is the centre of
the field of view (which is a point not imaged by the detectors).  The
remaining columns give the total integration time $t_{\rm tot}$ of the
frames contributing to the stack (i.e. excluding deprecated frames),
and the $5\sigma$ depth, for the {\em J} and {\em K} frames.

Fig. \ref{dxsarea} illustrates the distribution of depths of the DXS
observations in the DR1+ database, plotting the summed area that
attains a given minimum depth.

\subsubsection{Ultra Deep Survey}

\begin{table}
\centering
\begin{tabular}{|c|c|c|c|} 
\hline
 Band & $t_{\rm tot} $ & Depth & seeing \\
      &   s     & 5$\sigma$ & $\arcsec$ \\
 \hline
 $J$ & 21060 & 22.61 & 0.86 \\
 $K$ & 33910 & 21.55 & 0.76 \\
\hline  
\end{tabular}
\caption{UDS deep stacks. The table lists the total integration time
  in seconds, the 5$\sigma$ point source depth, and the image FWHM, in
  the {\em J} and {\em K} bands.}
\label{tab_uds_deep_stacks}
\end{table}

As already noted, for the UDS the EDR included data taken up to 2005
September 27. The DR1 includes additional observations, and the new
stacked images reach substantially deeper. The total integration times
on source, and depths reached in $J$ and $K$, and the seeing, are summarised in
Table \ref{tab_uds_deep_stacks}. As explained in Section \ref{update}
the UDS data are mosaiced. The individual $3\times3$ microstepped
detector frames are resampled onto a $\sim 24000 \times 24000$ pixel
grid, and the data are then stacked. In the WSA, the final frame
is referred to as a {\em mosaicdeepleavstack} frame.  Unlike
all other stacked data in the WSA, astrometry of the deep UDS stacks
uses tangential projection (TAN) rather than the zenith polynomial
projection (ZPN) that is standard for WFCAM.

\subsection{Survey status summary}

We have estimated the percentage completion of the surveys relative to
the final 7-year goals set out in \citet{lawrence06}, by computing the
product over all filters of the area covered and the summed effective
integration times (established from the depth achieved), in the DR1+
database. Computed in this way the completeness of the surveys is as
follows: LAS 6 per cent, GPS 7 per cent, GCS 13 per cent, DXS 11 per
cent, UDS 4 per cent. Overall, DR1 marks completion of 7 per cent of
the UKIDSS final goals.

\vspace{5mm}
\begin{flushleft}
{\bf Acknowledgements}
\end{flushleft}

We are grateful to Roc Cutri for several helpful discussions.

\setlength{\bibhang}{2.0em}

\end{document}